\documentclass[12pt]{article}
\usepackage{amsmath}
\usepackage{amssymb}
\usepackage{graphicx}
\def \be{\begin{equation}}
\def \ee{\end{equation}}
\def \bea{\begin{eqnarray}}
\def \eea{\end{eqnarray}}
\def \p{\partial}
\def \nn{\nonumber}

\def \G{\Gamma}
\def \o{\omega}
\def \O{\Omega}

\begin{document}
\title {Matrix Model in a Class of Time Dependent Supersymmetric Backgrounds}

\vspace{2cm}
\author{Hong-Zhi Chen\thanks{Email:hzhchen@pku.edu.cn},\hspace{5ex}Bin Chen\thanks{Email:bchen01@pku.edu.cn} \\
\vspace{2cm} School of Physics, Peking University, Beijing 100871,
P. R. China}
\date{}
\maketitle

\begin{abstract}
We discuss the matrix model in  a class of 11D time dependent
supersymmetric backgrounds as obtained in \cite{bchen}. We
construct the matrix model action through the matrix
regularization of the membrane action in the background. We show
that the action is exact to all order of fermionic coordinates.
Furthermore We discuss the fuzzy sphere solutions in this
background.
\end{abstract}
\newpage

\section{Introduction}

It is very important to understand string theory in the time
dependent background because this issue is related to some
fundamental questions in quantum gravity. One question is the
resolution of the cosmological singularities. Near the big-bang or
big-crunch singularity, the quantum effects should play an important
role and a quantum gravity description is needed. However, despite
of many efforts in the past decades, we are still far from
understanding the issue clearly. To address the issue, we have to
decide what the right degrees of freedom are to describe the physics
there. If the string coupling is small, we might hope that the
perturbative string is suitable. One class of models, called null
orbifold, have been constructed to investigate this possibility
\cite{nullorbifold}. These models keep part of the supersymmetries
and are solvable perturbatively. Unfortunately, it turned out that
these time dependent orbifold models are unstable to large back
reaction because of blue shifting of modes in these
background\cite{HP}. Quite recently, E. Silvertein et al propose
that a closed string tachyon condensate smooths out the singularity
by consistently massing up the degrees of freedom of the
system\cite{McGreevy:2005ci}.

If the string coupling is large near the singularity, one must
take the nonpurturbative string effects seriously. Very recently,
the authors in \cite{craps} raised the idea of matrix big bang.
They considered a type IIA theory in a null linear dilaton
background which preserves one-half of the original supersymmetry.
In this time-dependent background, the string coupling becomes
large near the big-bang singularity. In \cite{craps}, the authors
proposed a dual matrix string which is a two-dimensional
Super-Yang-Mills theory on the Milne orbifold to describe the big
bang where the Yang-Mills coupling becomes weak. In short, the
matrix degrees of freedom, rather than the point particle or the
perturbative string, describe the physics near big-bang
singularity. Following \cite{craps}, many authors discussed the
generalization of their background \cite{mli05}-\cite{lin06}.

In \cite{bchen}, a large class of rather general time-dependent
configurations  have been found in M-Theory. These configurations
keep sixteen supersymmetries, with killing spinor satisfying
$\Gamma^+\epsilon=0$, and has a null Killing field. Moreover, it
has been proved that such configurations generally have no
supernumerary supersymmetries. As a consequence, the corresponding
matrix model constructed by the DCLQ prescription has no linearly
realized supersymmetry.

One subtle point  in \cite{bchen} is that the construction of the
matrix model follows the route of the weak field approximation
\cite{taylor}. However, in the early time the configurations turn
out to be far from flat. It seems that the matrix model
construction in \cite{bchen} is in doubt. Generically the matrix
model in curved backgrounds is a subtle issue. Besides the weak
field approximation, there is another way to get the matrix model
action. The way is to start from the supermembrane action embedded
in a 11D curved background, and then transfer to the matrix model
action through matrix regularization. This has been proved
successful in recovering the BFSS matrix model from the membrane
in the flat spacetime \cite{BFSS,dwit88} and BMN matrix model from
the membrane in the 11D maximally supersymmetric plane-wave
background \cite{BMN,Dasgupta02}. However it should be noticed
that in a general curved background, the membrane action could
only be obtained order by order of fermionic coordinates $\theta$.
Up to order of $\theta^2$, the explicit form of the action has
been worked out in \cite{dwit98}.

In this short note, we would like to construct the matrix model
action of the configurations found in \cite{bchen}, following the
route of matrix regularization of supermembrane action. In our case,
we manage to get the exact membrane action to all order of $\theta$.
Moreover, we will discuss the evolution of the fuzzy sphere
solutions of the model. We will find that the radius of the fuzzy
sphere shrinks to zero in a big-bang like evolution while it grows
without limit in a big crunch like evolution.

\section{The Matrix Model}

Let us first give a short review of the background.  The metric of
our background is as follows:
\begin{eqnarray}\label{metric}
ds^2 &=& 2e^{r_0u}dudv+\sum\limits_ic_ie^{r_iu}(x^i)^2(du)^2+\sum\limits_ie^{r_iu}(dx^i)^2\nonumber\\
&+& \sum\limits_{ij}A^0_{ij}e^{(r_i+r_j)u/2}x^jdx^idu,
\end{eqnarray}
where
\begin{equation}
A^0_{ij} = -A^0_{ji}=const,
\end{equation}
and $r_0, r_i$ are all constants,too. We also have a four-form field
strength
\begin{equation}\label{field}
F_{u123} =e^{(r_1+r_2+r_3)u/2}f^0, \hspace{3ex}f^0=const.
\end{equation}
Our convention is as follows: we use $x^\mu$ for the curved space
coordinates with
\begin{equation}
x^\mu=(x^u, x^v, x^i)\equiv(u, v, x^i),
\end{equation}
where $u\equiv x^u=(x^{\mu=10}+x^{\mu=0})/\sqrt{2}$, and $v\equiv
x^v=(x^{\mu=10}-x^{\mu=0})/\sqrt{2}$, and $i=(1, \cdots , 9)$.
Similarly, We use $x^r$ to represent tangent space coordinates with
\begin{equation}
x^r=(x^+, x^-, x^I),
\end{equation}
where $x^+=(x^{r=10}+x^{r=0})/\sqrt{2}$, and
$x^-=(x^{r=10}-x^{r=0})/\sqrt{2}$, and $I=(1, \cdots, 9)$.

The background keeps sixteen ``standard" supersymmetries
characterized by Killing spinor satisfying $\Gamma^+\epsilon=0$.
There is no supernumerary supersymmetry in this case. This indicates
that there is no linearly realized supersymmetries in the embedded
supermembrane action and hence in the matrix model action. Another
remarkable fact is that there exist a null Killing vector in the
background. And also the Ricci tensor and the field strength have no
lower index in $v$ and no dependence on $v$.

 We will
begin to derive the matrix model in this background following
\cite{dwit98}. The supermembrane action is:
\begin{equation}
S[Z(\xi)]=\int d^3\xi
[-\sqrt{-g(Z(\xi))}-\frac{1}{6}\epsilon^{abc}\Pi^A_a\Pi^B_b\Pi^C_cB_{CBA}(Z(\xi))],
\end{equation}
where $Z^A(\xi)=(x^\mu(\xi),\theta(\xi))$ is the curved superspace
coordinates, $g_{ab}=\Pi^ {\mu}_a\Pi^{\nu}_b g_{\mu
\nu}=\Pi^r_a\Pi^s_b\eta_{rs}$ is the induced metric,
$\eta_{rs}=diag(-1,1,...1)$ is the 11-d Lorentz metric, and
$\xi^a=(\xi^0, \xi^1, \xi^2)=(\tau, \xi^{\alpha})$, $\alpha=1,2$
represent the coordinates on the world volume. Here $\Pi^ A_a$ are
the supervielbein pullback, $B_{ABC}$ are the super three-potential.
In \cite{dwit98}, the authors have obtained the expression of these
two quantities in terms of component fields to order $\theta^2$ of
fermionic coordinates. In our case, the gravitino is zero, so the
supervielbein pullback is:
\begin{eqnarray}\label {pi}
\Pi^r_a &=& \partial_aZ^AE^r_A \nonumber \\
&=& \partial_ax^\mu(e^r_\mu-\frac{1}{4}{\bar
\theta}\Gamma^{rst}\theta\omega_{\mu st}+{\bar
\theta}\Gamma^r\Omega_\mu\theta)+{\bar
\theta}\Gamma^r\partial_a\theta+{\mathcal O}(\theta^3),
\end{eqnarray}
where $\omega_{\mu st}$ is the spin connection, and
\begin{equation}
\Omega_\mu=\frac{1}{288}F_{\nu\rho\sigma\lambda}(\Gamma^{\nu\rho\sigma\lambda}_\mu
+8\Gamma^{\nu\rho\sigma}\delta^\lambda_\mu).
\end{equation}
The super three-potential pullback is:
\begin{eqnarray}\label {potential}
&-&\frac{1}{6}\Pi^A_a\Pi^B_b\Pi^C_cB_{CBA} \nonumber \\
&=&\frac{1}{6}\epsilon^{abc}\partial_a x^\mu\partial_b
x^\nu\partial_c x^\rho\Big[C_{\mu\nu\rho} +\frac{3}{4}{\bar
\theta}\Gamma_{rs}\Gamma_{\mu\nu}\theta\omega^{rs}_\rho -3{\bar
\theta}\Gamma_{\mu\nu}\Omega_\rho\theta\Big] \nonumber \\
&-&\epsilon^{abc}{\bar
\theta}\Gamma_{\mu\nu}\partial_c\theta\Big[\frac{1}{2}\partial_ax^\mu(\partial_bx^\nu+{\bar
\theta}\Gamma^\nu\partial_b\theta)
+\frac{1}{6}{\bar\theta}\Gamma^\mu\partial_a\theta{\bar\theta}\Gamma^\nu\partial_b\theta\Big]+{\mathcal
O}(\theta^3),
\end{eqnarray}
where $C_{\mu\nu\rho}$ is the three-form potential. To simplify the
action, we go to light-cone gauge:
\begin{equation}
x^u=u=\tau.
\end{equation}
And because of the $\kappa$-symmetry of the action, we can also
impose an additional gauge \cite{dwit98}
\begin{equation}
\Gamma^+\theta=0.
\end{equation}
We further decompose our 11D gamma matrices $\Gamma^I$ using 9D
matrices $\gamma^I$ as follows:
\begin{eqnarray}
\Gamma^I &=& \gamma^I\otimes \sigma_3, (I=1,...,9), \\
\Gamma^0 &=& 1\otimes i\sigma_1, \\
\Gamma^{11} &=& -1\otimes \sigma_2, \\
\Gamma_- &=& \Gamma^+=\frac{1}{\sqrt{2}}(\Gamma^0+\Gamma^{11}), \\
\Gamma_+ &=& \Gamma^-=\frac{1}{\sqrt{2}}(-\Gamma^0+\Gamma^{11}).
\end{eqnarray}
Then $\theta$ can be decomposed as:
\begin{eqnarray}
\theta &=& \frac{1}{2^{1/4}}(\psi^T, 0)^T, \\
\bar \theta &=& \frac{1}{2^{1/4}}(0, -\psi^T).
\end{eqnarray}

Using the formula of \cite{dwit98}, we can only derive the
following formulae up to ${\mathcal O}(\theta^2)$. But as we will
argue at the end of this section, our matrix model is exact to all
orders of $\theta$. So we omit the terms higher than $\theta^2$ in
the following formulae.

Plugging the above expressions into the action, for the metric we
get
\begin{eqnarray}
g_{\alpha\beta} &=& \sum\limits_{i}e^{r_i\tau}\partial_\alpha x^i\partial_\beta x^i,  \\
g_{00} &=&
\sum\limits_ic_ie^{r_i\tau}(x^i)^2-\sum\limits_{IJ}\frac{i}{4}e^{r_0\tau/2}A^0_{IJ}\psi^T\gamma^{IJ}\psi
-\frac{i}{3}e^{r_0\tau/2}f^0\psi^T\gamma^{123}\psi \nonumber\\
 &+& 2e^{r_0\tau}\partial_0v+2ie^{r_0\tau/2}\psi^T\partial_0\psi+
 \sum\limits_{ij}A^0_{ij}e^{(r_i+r_j)\tau/2}x^j\partial_0x^i \nonumber\\
  &+& \sum\limits_ie^{r_i\tau}(\partial_0x^i)^2.
\end{eqnarray}
We don't need the explicit form of $u_\alpha\equiv g_{0\alpha}$ in
the later calculations, and we only need to know that it depends on
$\dot X^i$. For the super three-form potential term, we get
\begin{eqnarray}
&-& \frac{1}{6}\epsilon^{abc}\Pi^A_a\Pi^B_b\Pi^C_cB_{CBA}(Z(\xi)) \nonumber\\
&=&
-i\sum\limits_{I,i}\psi^T\gamma^I\{x^i,\psi\}e^{(r_0+r_i)\tau/2}\delta^I_i\nonumber \\
&-&\frac{1}{2}\sum\limits_{i,j=1,2,k=3}\{x^i,x^j\}x^k\epsilon_{ijk}f^0e^{(r_i+r_j+r_k)\tau/2},
\end{eqnarray}
where
\begin{equation}
\{A,B\}=\epsilon^{\alpha\beta}\partial_\alpha A\partial_\beta B.
\end{equation}
Now, we decompose $g=det(g_{ab})$ as following:
\begin{equation}
g=-\Delta\bar g,
\end{equation}
where
\begin{eqnarray}
\bar g_{\alpha\beta} &=& g_{\alpha\beta},  \\
\bar g &=& det(\bar g_{\alpha\beta}),  \\
\bar g^{\alpha\beta}\bar g_{\beta\gamma} &=& \delta^\alpha_\gamma, \\
\Delta &=& -g_{00}+u_\alpha\bar g^{\alpha\beta}u_\beta.
\end{eqnarray}
To solve the constraints, we go to the Hamiltonian formalism. We get
the expression for the canonical momentum of the $X^i, v$, and
$\psi$:
\begin{eqnarray}
P_v &=& P^u=e^{r_0\tau}\sqrt{\frac{\bar g}{\Delta}}, \\
P_\psi &=& ie^{r_0\tau/2}\sqrt{\frac{\bar g}{\Delta}} \psi^T=ie^{-r_0\tau/2}\psi^TP^u, \\
P_i &=& \sqrt{\frac{\bar g}{\Delta}}\big(e^{r_i\tau}\partial_0
x^i+\frac{1}{2}\sum\limits_j
A^0_{ij}e^{(r_i+r_j)\tau/2}x^j-e^{r_i\tau}\partial_\alpha x^i\bar
g^{\alpha\beta}u_\beta\big).
\end{eqnarray}
After the Legendre transformation:
\begin{equation}
{\cal H}=\sum\limits_iP_i\dot x^i+P_v\dot v+P_\psi\dot\psi-{\cal L},
\end{equation}
we have the hamiltonian density:
\begin{eqnarray}
{\cal H} &=& \sum\limits_i\frac{P^2_i}{2p^\tau}e^{(r_0-r_i)\tau}+
\frac{e^{r_0\tau}}{4P^u}\sum\limits_{ij}e^{(r_i+r_j)\tau}\{x^i,x^j\}^2\nonumber\\
&-& \frac{1}{2}\sum\limits_{ij}A^0_{ij}e^{(r_j-r_i)\tau/2}x^jP_i
+i\sum\limits_{I,i}\psi^T\gamma^I\{x^i,\psi\}e^{(r_0+r_i)\tau/2}\delta^I_i \nonumber \\
&+&\frac{1}{2}\sum\limits_{i,j=1,2,k=3}\{x^i,x^j\}x^k\epsilon_{ijk}f^0e^{(r_i+r_j+r_k)\tau/2}
\nonumber\\
&+&
\frac{1}{2}e^{-r_0\tau}P^u\big[-\sum\limits_ic_ie^{r_i\tau}(x^i)^2
+\frac{i}{4}\sum\limits_{IJ}e^{r_0\tau/2}A^0_{IJ}\psi^T\gamma^{IJ}\psi\nonumber\\
&+& \frac{i}{3}e^{r_0\tau/2}f^0\psi^T\gamma^{123}\psi
+\frac{1}{4}\sum\limits_{ijk}A^0_{ij}A^0_{ik}e^{(r_j+r_k)\tau/2}x^jx^k\big].
\end{eqnarray}
This Hamiltonian density can be derived from a Lagrangian density
consisting of only physical degrees of freedom $x^i$ and $\psi$ of
the following form:
\begin{eqnarray}
{\cal L} &=&\sum\limits_i\frac{P^u}{2}e^{(r_i-r_0)\tau}(D_\tau
x^i)^2
+\frac{P^u}{2}\sum\limits_{ij}A^0_{ij}e^{(\frac{r_i+r_j}{2}-r_0)\tau}x^jD_\tau x^i\nonumber\\
&+& \frac{P^u}{2}\sum\limits_ic_ie^{(r_i-r_0)\tau}(x^i)^2
-\frac{e^{r_0\tau}}{4P^u}\sum\limits_{ij}e^{(r_i+r_j)\tau}\{x^i,x^j\}^2\nonumber\\
&-&\frac{1}{2}\sum\limits_{i,j=1,2,k=3}\{x^i,x^j\}x^k\epsilon_{ijk}f^0e^{(r_i+r_j+r_k)\tau/2}\nonumber\\
&+& iP^ue^{-r_0\tau/2}\psi^T D_\tau\psi
-\frac{i}{6}P^ue^{-r_0\tau/2}f^0\psi^T\gamma^{123}\psi \nonumber\\
&-&\frac{i}{8}P^u\sum\limits_{IJ}e^{-r_0\tau/2}A^0_{IJ}\psi^T\gamma^{IJ}\psi
-i\sum\limits_{I,i}\psi^T\gamma^I\{x^i,\psi\}e^{(r_0+r_i)\tau/2}\delta^I_i,
\end{eqnarray}
where $D_\tau$ is the covariant derivative with respect to an
auxiliary gauge field $A_0$.

Now, let us do the usual matrix regularization:
\begin{eqnarray}
x^i \rightarrow X^i_{N\times N}, \\
\psi \rightarrow \psi_{N\times N}, \\
P^u \int d^2 \sigma \rightarrow \frac{1}{R} Tr,\\
\{,\} \rightarrow -i[,],
\end{eqnarray}
in the above membrane action, and we finally obtain the matrix
model action:
\begin{eqnarray}\label{action}
S &=& \int d\tau Tr \Big(
\sum\limits_i\frac{1}{2R}e^{(r_i-r_0)\tau}(D_\tau X^i)^2
+\frac{1}{2R}\sum\limits_{ij}A^0_{ij}e^{(\frac{r_i+r_j}{2}-r_0)\tau}X^j D_\tau X^i\nonumber\\
&+& \frac{1}{2R}\sum\limits_ic_ie^{(r_i-r_0)\tau}(X^i)^2
+\frac{R}{4}e^{r_0\tau} \sum\limits_{ij}e^{(r_i+r_j)\tau}\big[X^i,X^j\big]^2\nonumber\\
&+&\frac{i}{2}\sum\limits_{i,j=1,2,k=3}\big[X^i,X^j\big]X^k\epsilon_{ijk}f^0e^{(r_i+r_j+r_k)\tau/2}\nonumber\\
&+& \frac{i}{R}e^{-r_0\tau/2}\psi^T D_\tau\psi
-\frac{i}{6R}e^{-r_0\tau/2}f^0\psi^T\gamma^{123}\psi \nonumber\\
&-&\frac{i}{8R}\sum\limits_{IJ}e^{-r_0\tau/2}A^0_{IJ}\psi^T\gamma^{IJ}\psi
-\sum\limits_{I,i}\psi^T\gamma^I\big[X^i,\psi\big]e^{(r_0+r_i)\tau/2}\delta^I_i\Big).
\end{eqnarray}
Although this action seems a bit cluttered, it can be cast into a
canonical form by rescalings \cite{das03}. The bosonic part of the
action is the same as the one raised in \cite{bchen}\footnote{While
completing the manuscript, we realized that in \cite{bchen}, there
exists two typos in the bosonic action. One is a sign difference,
the other is due to the overcounting of the background 3-form
potential. After fixing them, the bosonic action in \cite{bchen}
agrees with the above one.}, while the fermionic part is different
by the prefactors. The main discrepancy in the fermionic actions
comes from the $g^{uv}$ factor. We suspect that such factors have
not been incorporated properly in the weak field approximation. We
believe that the treatment in this paper is more convincing.

Although we have derived this matrix model using formulae of
\cite{dwit98} that are only exact to order $\theta^2$, we will now
argue that it is in fact exact to all orders of $\theta$. The
argument is quite similar to that in \cite{iizuka}, in which the
authors argued that his matrix model on a pp-wave background is
exact to all orders of $\theta$. Similar argument has been used in
the discussion of the Green-Schwarz string action in a class of
plane-wave background \cite{Tseytlin02}. The main points are as
follows. First notice that the supervielbein pullback
$\Pi^r_a=\partial_aZ^AE^r_A$ is linear in $\partial_a X^\mu$, while
$E^r_A$ is constituted with other quantities. It can be seen from
their explicit form that these other quantities, $\theta$,
$\Gamma^r$, Ricci tensor, $\Omega_\mu$, and field strength et.al.
have no lower curved spacetime index $v$, and hence no upper curved
spacetime index $u$. Also from the form of the metric, we notice
that the only spin connections $\omega_{\mu\nu\rho}=\omega^{rs}_\mu
e_{r\mu}e_{s\nu}$ with lower curved spacetime index $v$ is
$\omega_{uuv}$, and the only geometrical object with lower index $v$
constructed from the vielbein $e^r_\mu e_{r\nu}$ and their
derivatives must have the lower index $uv$ appearing at the same
time. Hence, althoug these two quantities can have upper curved
spacetime index $u$, they must also have lower curved spacetime
index $u$ at the same time. On the other hand, the nonvanishing
bilinear fermionic terms $\bar{\theta}\Gamma^{rst\cdots}\theta$
always have one and only one $\Gamma^-$ and no $\Gamma^+$ due to the
gauge condition $\Gamma^+\theta=0$. The upper tangent space index
$r=-$ require an upper curved spacetime index $\mu=u$ coming from
other geometrical quantities because the only nonzero vielbein with
a lower tangent index $r=-$ is $e_{-u}$. Such an index cannot be
cancelled by the above mentioned quantities except $\partial_a X^u$.
For example, the other two possible quantities with the upper curved
index $u$ must carry the lower curved spacetime index $u$ at the
same time. So the net result is to leave a lower curved index $u$.
This index can only be cancelled by $\partial_aX^u$. But due to the
linearity in $\partial_aX^\mu$, one at most has bilinear $\theta$
terms in $\Pi_a^A$. Also as a consequence, the super three-potential
pullback term can only have bilinear $\theta$ terms. This is due to
the antisymmetric nature of $\epsilon^{abc}$ and the fact that
bilinear $\theta$ term in $\Pi^A_a$ must be proportional to
$\partial_aX^u$. In short, the expressions
(\ref{pi},\ref{potential}) have vanishing higher order terms and so
are exact to all order of $\theta$. Therefore the matrix model
(\ref{action}) is exact to all orders of fermionic coordinates.

\section{The Fuzzy Sphere Solution}

\begin{figure}
  \includegraphics[width=0.3\textwidth]{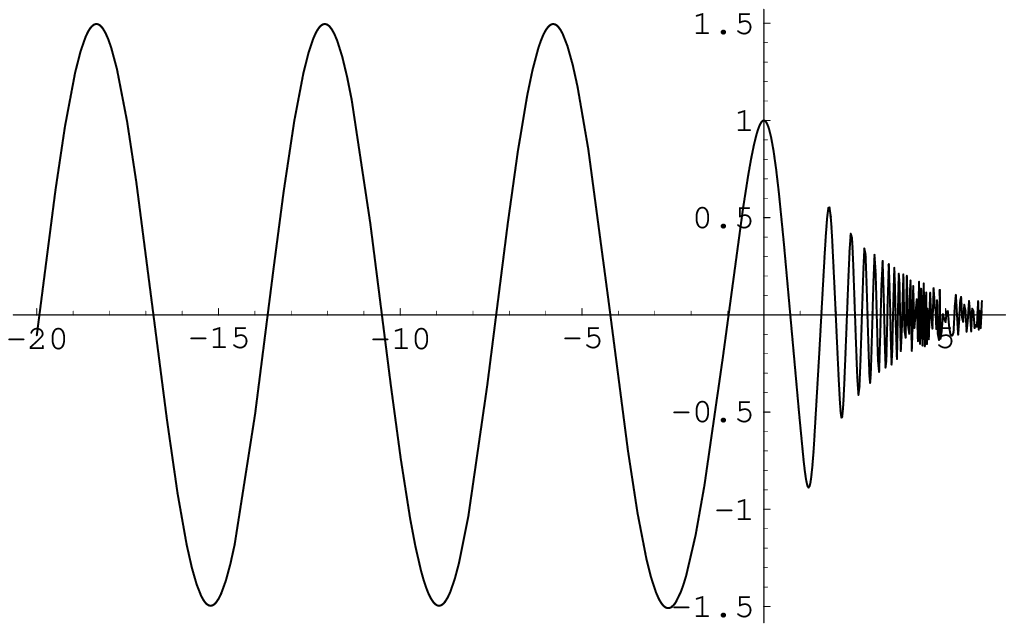}\hspace{0.5cm}
  \includegraphics[width=0.3\textwidth]{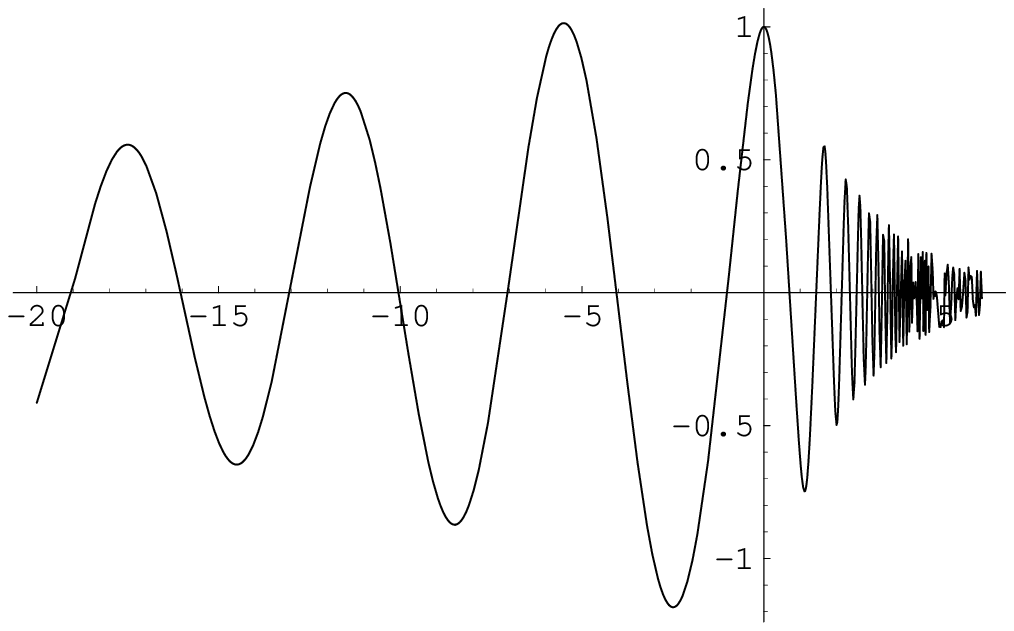}\hspace{0.5cm}
  \includegraphics[width=0.3\textwidth]{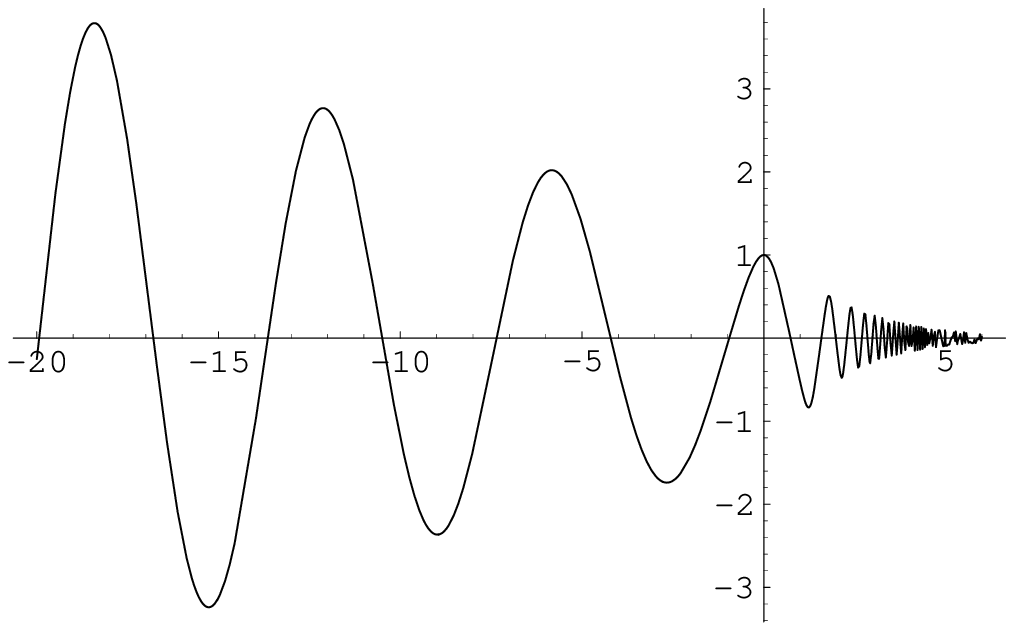}
  \caption{Fuzzy sphere solution with $\hat{r}=1, 0.8, 1.1$ respectively.}\label{positive}
\end{figure}

\begin{figure}
  \centerline{\includegraphics[width=0.5\textwidth]{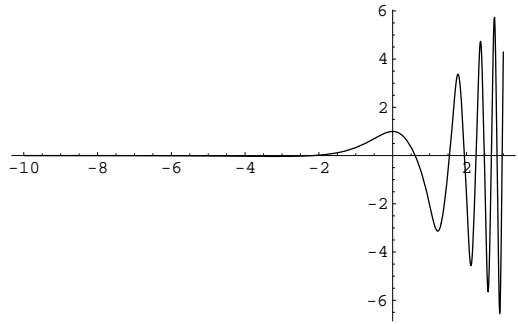}}
  \caption{Fuzzy sphere solution with $\hat{r}=-1$.}\label{negative}
\end{figure}

We would like to discuss the fuzzy sphere solution of the classical
equation of motion derived from the matrix model action. To
investigate the simplest situation, consider the matrix model in the
sector:
\begin{equation}
X^4=X^5=...=X^9=0, \psi=0.
\end{equation}
To further simplify the problem, we restrict ourselves to symmetric
case with:
\begin{equation}
r_1=r_2=r_3=r, c_1=c_2=c_3=c.
\end{equation}
We want to find solution of the form:
\begin{equation}
X^a(\tau)=S(\tau)J^a, a=1,2,3,
\end{equation}
where $J^a$ is N dimensional representation of $SU(2)$. Use
\begin{equation}
Tr\sum\limits_a(J^a)^2=\frac{N(N-1)}{4},
\end{equation}
and
\begin{equation}
[J^a,J^b]=i\epsilon^{abc}J^c.
\end{equation}
We finally get
\begin{equation}
\frac{d^2S}{d\tau^2}+(r-r_0)\frac{dS}{d\tau}
+2R^2e^{(2r_0+r)\tau}S^3+Rf^0e^{(r_0+r/2)\tau}S^2 -cS=0.
\end{equation}
We change this equation to be dimensionless by introducing two
dimensionless variables:
\begin{equation}
t=r_0\tau, S(\tau)=R\tilde{S}(t),
\end{equation}
We insert appropriate powers of $l_p$ and further introduce the
other dimensionless variables as follows:
\begin{equation}
\hat{r}=\frac{r}{r_0},\hat{c}=\frac{c}{r^2_0},\hat{f}^0=\frac{f^0}{2\sqrt{2}r_0},A=\frac{\sqrt{2}R^2}{r_0l^3_p}.
\end{equation}
The equation then becomes:
\begin{equation}
\frac{d^2\tilde{S}}{dt^2}+(\hat{r}-1)\frac{d\tilde{S}}{dt}
+A^2e^{(2+\hat{r})t}\tilde{S}^3+2\hat{f}^0Ae^{(1+\hat{r}/2)t}\tilde{S}^2-\hat{c}\hat{S}=0.
\end{equation}
To investigate the behavior of fuzzy sphere solution as the time
evolves, we have numerically solved this equation. We chose initial
conditions as:
\begin{equation}
\tilde{S}(0)=1, \tilde{S} '(0)=0,
\end{equation}
and choose the dimensionless parameters as:
\begin{equation}
\hat{c}=-1, \hat{f}^0=1, A=1.
\end{equation}
The behavior for $\hat{r}=1, 0.8, 1.1$ respectively is shown in
figure \ref{positive}.
We see that the behavior is similar to that of \cite{das}, i.e. the
radius of the fuzzy sphere shrinks to zero at late times as the
spatial dimensions expand larger and larger in a big-bang like
evolution. This is expected, as when the spatial dimensions expand
larger, the effect of non-Abelian degrees of freedom of the matrix
model become less important. The above is for the case $\hat{r}>0$,
for the case $\hat{r}<0$, the evolution is different. In this case,
the evolution is big-crunch like and we expect that as the spatial
dimensions collapse, the effect of the non-Abelian degrees of
freedom of the matrix model will become more and more important. We
see this behavior in figure \ref{negative} that the radius of the
fuzzy sphere grows as time evolves.

The above behaviors at late time are also seen in the cases with
other parameters choosing different values as far as they do not
change their signs. Also the behaviors are the same if we change the
initial condition. We also investigated the more general cases with
non-symmetric metric, i.e. with different $c_i$'s and different
$r_i$'s. Again, the late time behaviors are the same. So we conclude
that our fuzzy sphere solutions are reasonable, and that the matrix
model we derived is also reasonable.

\section{Conclusions and Discussions}

In this note, we studied the matrix model action in a class of the
supersymmetric time-dependent backgrounds. We first discussed the
membrane action in the background and then through matrix
regularization we obtained the corresponding matrix model action.
One remarkable fact is that our background, though slightly
different from the plane-wave background, still permits us to get
the exact action to all orders of fermionic coordinates. This fact
shows that although these configurations does not keep full
supersymmetry, they are easier to deal with than the ordinary
curved spacetime.  It would be nice to investigate these
configurations and their matrix models more thoroughly. In this
paper, we studied some fuzzy sphere like classical solution and
found they share the same property uncovered in \cite{das}. It
would be interesting to study 1-loop \cite{mli0512,crap0601,
martinec06}, brane creation\cite{das0602} issues in these
backgrounds.

Our discussion focused on the configurations
(\ref{metric},\ref{field}), which is a special class of the general
supersymmetric time-dependent configurations (\ref{metric3},\ref{F})
in the appendix. The study of the above sections can be generalized
to the general backgrounds straightforwardly. Especially, the
argument of the exactness still make sense. This can be seen from
the explicit form of the metric, orthogonal frame, spin connections
and field strength. This means that the matrix model action in the
configurations (\ref{metric3},\ref{F}) would be exact to all order
of the fermionic coordinates. It deserves more study.

Our construction shed some light on the relation between membrane
regularization method and usual DCLQ prescription to construct the
matrix model action. It turns out that the membrane regularization
method is quite effective. It should be straightforward to
generalize the method to the construction of matrix string action.
In \cite{das}, it has been shown the equations of motion of the
membrane and the fuzzy sphere is the same. This suggests that the
matrix string action there could be obtained by matrix
regularization of membrane action.

\section*{Acknowledgements}

We would like to thank F.L. Lin for the discussion which inspired
this project. The work was supported by NSFC Grant No.
10405028,10535060 and the Key Grant Project of Chinese Ministry of
Education (NO. 305001)

\section*{Appendix}

In this appendix, we collect some relations on the configurations
discussed in \cite{bchen}. The general supersymmetric
time-dependent backgrounds have a metric of form
\be
\label{metric3}
 ds^2=2A_0(u)dudv+B_{ij}(u)x^ix^j(du)^2+A_i(u)(dx^i)^2+A_{ij}(u)x^jdx^idu,
 \ee
with $B_{ij}(u)=B_{ji}(u)$ and $A_{ij}(u) = -A_{ji}(u)$, and have
the field strength \be \label{F}
 F_{u123}=f_0(u).
 \ee
The metric (\ref{metric3}) allows an orthogonal frame
 \bea
 e^+&=&\sqrt{A_0(u)}du \\
 e^-&=&\sqrt{A_0(u)}dv+\frac{B_{ij}(u)x^ix^j}{2\sqrt{A_0(u)}}du+\frac{A_{ij}(u)x^j}{2\sqrt{A_0(u)}}dx^i
 \\
 e^I&=&\sqrt{A_i(u)}dx^i\delta^I_i.
 \eea
The corresponding spin connections are
 \bea
 \o^{-+}&=&-\frac{\p_u\sqrt{A_0}}{\sqrt{A_0}}du \nn\\
 \o^{+i}&=&0 \nn \\
 \o^{ij}&=&-\frac{A_{ji}}{2\sqrt{A_iA_j}}du \nn \\
 \o^{-i}&=&\frac{1}{\sqrt{A_i}}\left(\frac{B_{ij}x^j}{\sqrt{A_0}}-\frac{\p_uA_{ij}x^j}{2\sqrt{A_0}}
 +\frac{\p_u\sqrt{A_0}}{A_0}A_{ij}x^j\right)du\nn\\
 & &-\frac{\p_u\sqrt{A_i}}{\sqrt{A_0}}dx^i+\sum_{j\neq i}\frac{A_{ji}}{2\sqrt{A_0A_i}}dx^j.
  \eea
With the field strength, we have \bea
 \O_v&=&0 \\
 \O_u&=&-\frac{1}{12}(\G^{+-123}+\G^{123})\frac{f_0}{\sqrt{A_1A_2A_3}}
 \\
 \O_i&=&\frac{1}{24}(3\G^{123}\G^i+\G^i\G^{123})\G^+\frac{\sqrt{A_i}f_0}{\sqrt{A_0A_1A_2A_3}}.
\eea

The Ricci tensor has the only
 nonvanishing component
 \be
 R_{uu}=\sum_i\frac{\sqrt{A_0}}{\sqrt{A_i}}\left(-\p_u(\frac{\p_u\sqrt{A_i}}{\sqrt{A_0}})-
 \frac{1}{\sqrt{A_0A_i}}B_{ii}+
  \frac{\p_u\sqrt{A_0}\p_u\sqrt{A_i}}{A_0}+\sum_{j\neq
  i}\frac{A^2_{ij}}{4A_j\sqrt{A_0A_i}}\right).
  \ee
The nontrivial equation of motion is
 \be
 R_{uu}=\frac{f_0^2}{2A_1A_2A_3}
 \ee

\end{document}